\documentstyle[11pt,moriond]{article}
\input{psfig}

\def\Journal#1#2#3#4{{#1} {\bf #2}, #3 (#4)}

\def\NPB{{\em Nucl. Phys.} B}
\def\PLB{{\em Phys. Lett.}  B}
\def\PRL{\em Phys. Rev. Lett.}
\def\PRD{{\em Phys. Rev.} D}

\def\APJ{{\em Astrophys. J.}}
\def\APJL{{\em Astrophys. J. Lett.}}

\def\MNRASL{{\em Mon. Not. R. Astron. Soc.}}

\def\be{\begin{equation}}
\def\ee{\end{equation}}
\def\bea{\begin{eqnarray}}
\def\eea{\end{eqnarray}}

\def\VEV#1{\left\langle #1\right\rangle}
\def\sec{\ifmmode \,\, {\rm sec} \else sec \fi}
\def\eV {\ifmmode \,\, {\rm eV} \else eV \fi}
\def\keV{\ifmmode \,\, {\rm keV} \else keV \fi}
\def\MeV{\ifmmode \,\, {\rm MeV} \else MeV \fi}
\def\GeV{\ifmmode \,\, {\rm GeV} \else GeV \fi}
\def\TeV{\ifmmode \,\, {\rm TeV} \else TeV \fi}
\def\fm{\ifmmode \,\, {\rm fm} \else TeV \fi}
\def\pbarn{\ifmmode \,\, {\rm pb} \else pb \fi}
\def\km{\ifmmode {\rm km}\, \else km \fi}
\def\Mpc{\ifmmode {\rm Mpc}\, \else Mpc \fi}
\def\Gyr{\ifmmode {\rm Gyr}\, \else Gyr \fi}

\def\fun#1#2{\lower3.6pt\vbox{\baselineskip0pt\lineskip.9pt
  \ialign{$\mathsurround=0pt#1\hfil##\hfil$\crcr#2\crcr\sim\crcr}}}
\def\la{\mathrel{\mathpalette\fun <}}
\def\ga{\mathrel{\mathpalette\fun >}}

\def\sbar#1{\kern 0.8pt
        \overline{\kern -0.8pt #1 \kern -0.8pt}
        \kern 0.8pt}  

\def\meter{\ifmmode \,\, {\rm m} \else m \fi}
\def\yr {\ifmmode \,\, {\rm yr} \else yr \fi}

\def\sr{\ifmmode \,\, {\rm sr} \else sr \fi}

\def\hatn{{\bf \hat n}}

\begin{document}
\vspace*{4cm}
\title{COSMOLOGICAL-PARAMETER DETERMINATION WITH COSMIC
MICROWAVE BACKGROUND TEMPERATURE ANISOTROPIES AND POLARIZATION}

\author{M. KAMIONKOWSKI}

\address{Department of Physics, Columbia University, 538 West
120th St., New York, NY 10027~~U.S.A}

\maketitle\abstracts{
Forthcoming cosmic microwave background experiments (CMB) will
provide precise new tests of structure-formation theories.  The
geometry of the Universe may be determined robustly, and the
classical cosmological parameters, such as the cosmological
constant, baryon density, and Hubble constant, may be determined
as well.  In addition, the ``inflationary observables,'' which
parameterize the shapes and amplitudes of the primordial spectra
of density perturbations and long-wavelength gravitational waves
produced by inflation, may also be measured and thus provide
several new tests of inflation.  Although most attention has
focussed on the more easily observed temperature anisotropies,
recent work has shown that the CMB polarization provides a
wealth of unique information that may be especially important
for determination of the inflationary observables.  Secondary
anisotropies at small angular scales produced by re-scattering
of photons from partial reionization may be used to constrain
the ionization history of the Universe.
}

\section{Introduction}

Despite its major triumphs (the expansion, nucleosynthesis,
and the cosmic microwave background), the big-bang theory for
the origin of the Universe leaves several questions unanswered.
Chief amongst these is the horizon problem:  When cosmic
microwave background (CMB) photons last scattered, the age of
the Universe was roughly 100,000 years, much smaller than its
current age of roughly 10 billion years.  After taking into
account the expansion of the Universe, one finds that the angle
subtended by a causally connected region at the surface of last
scatter is roughly $1^\circ$. However, there are 40,000 square
degrees on the surface of the sky.  Therefore, when we look at
the CMB over the entire sky, we are looking at 40,000
causally disconnected regions of the Universe.  But quite remarkably,
each has the same temperature to roughly one part in $10^5$!

The most satisfying (only?) explanation for this is slow-roll
inflation,\cite{inflation} a
period of accelerated expansion in the early Universe driven by
the vacuum energy most likely associated with a symmetric phase
of a GUT Higgs field (or perhaps Planck-scale physics or
Peccei-Quinn symmetry breaking).  Although the physics
responsible for inflation is still not well understood,
inflation generically predicts (1) a flat Universe; (2) that
primordial adiabatic (i.e., curvature) perturbations are
responsible for the large-scale structure (LSS) in the Universe
today;\cite{scalars} and (3) a stochastic gravity-wave
background.\cite{abbott}  More
precisely, inflation predicts a spectrum $P_s = A_s k^{n_s}$
(with $n_s$ near unity) of primordial density (scalar metric)
perturbations, and a stochastic gravity-wave background (tensor
metric perturbations) with spectrum $P_t =A_t \propto k^{n_t}$
(with $n_t$ small compared with unity).  (4) Inflation further
uniquely predicts specific relations between the
``inflationary observables,'' the amplitudes $A_s$ and $A_t$ and
spectral indices $n_s$ and $n_t$ of the scalar and tensor
perturbations.\cite{steinhardt} The amplitude of the
gravity-wave background is
proportional to the height of the inflaton potential, and the
spectral indices depend on the shape of the inflaton potential.
Therefore, determination of these parameters would illuminate
the physics responsible for inflation.  

Until recently, none of these predictions could really be tested.
Measured values for the density of the Universe span
almost an order of magnitude.  Furthermore, most do not probe
the possible contribution of a cosmological constant (or some
other diffuse matter component), so they do not address the
geometry of the Universe.  The only observable effects of a
stochastic gravity-wave background are in the CMB.  COBE
observations do in fact provide an upper limit to the tensor
amplitude, and therefore an inflaton-potential height near the
GUT scale.  However, there is no way to disentangle the scalar
and tensor contributions to the COBE anisotropy.

In recent years, it has become increasingly likely that
adiabatic perturbations are responsible for the
origin of structure.  Before COBE, there were numerous plausible
models for structure formation: e.g., isocurvature perturbations
both with and without cold dark matter, late-time or slow phase 
transitions, topological defects (cosmic strings or textures),
superconducting cosmic strings, explosive or seed models, a
``loitering'' Universe, etc.  However, after COBE, only
primordial adiabatic perturbations and topological defects were
still considered seriously.  And in the past few
months, some leading proponents of topological defects have
conceded that these models have difficulty reproducing the
observed large-scale structure.\cite{towel}

We are now entering an exciting new era, driven by new
theoretical ideas and developments in detector technology, in
which the predictions of inflation will be tested with
unprecedented precision.  It is even conceivable that early in
the next century, we will move from verification to
direct investigation of the high-energy physics responsible for
inflation.

The purpose of this talk is to review how forthcoming CMB
experiments will test several of these predictions.  I will
first review the predictions of inflation for density
perturbations and gravity waves.  I will then discuss how CMB
temperature anisotropies will test the inflationary predictions
of a flat Universe and a primordial spectrum of density
perturbations.  I review how a CMB polarization map
may be used to isolate the gravity waves and briefly review how
detection of these tensor modes may be used to learn about the
physics responsible for inflation.  I then discuss some recent
work on secondary anisotropies at smaller angular scales, and
how these may be used to probe the epoch at which objects first
underwent gravitational collapse in the Universe.  I close with
some brief remarks about further testable consequences of
inflation.

\section{Inflationary Observables}

Inflation occurs when the energy density of the Universe is
dominated by the vacuum energy $V(\phi)$ associated with some
scalar field $\phi$ (the ``inflaton'').  During this time, the
quantum fluctuations in $\phi$ produce classical scalar
perturbations, and quantum fluctuations in the spacetime metric
produce gravitational waves.  If the inflaton potential
$V(\phi)$ is given in units of $m_{\rm Pl}^4$, and the inflaton
$\phi$ is in units of $m_{\rm Pl}$, then the scalar and tensor
spectral indices are
\begin{equation}
     1-n_s = ( 1/ 8\pi) \left( V'/ V \right)^2 -
     (1/ 4 \pi) \left(V'/ V \right)', \qquad
     n_t = -( 1/ 8\pi) \left( V'/ V \right)^2. 
\label{spectralindices}
\end{equation}
The amplitudes can be fixed by their contribution to $C_2^{\rm TT}$,
the quadrupole moment of the CMB temperature,
${\cal S} \equiv  6\, C_2^{{\rm TT},{\rm scalar}}=
33.2\,[V^3/(V')^2]$, and ${\cal T} \equiv  6\, C_2^{{\rm
TT},{\rm tensor}}= 9.2 \,V.$
For the slow-roll conditions to be satisfied, we must have
$(1 /16 \pi) (V'/V)^2 \ll 1$, and $(1 /8\pi)(V''/V)  \ll  1$,
which guarantee that inflation lasts long enough to make the Universe
flat and to solve the horizon problem.

When combined with COBE results, current degree-scale--anisotropy and
large-scale-structure observations suggest that ${\cal T}/{\cal S}$ is less
than order unity in inflationary models, which restricts
$V\la 5\times 10^{-12}$.  Barring strange coincidences, the COBE
spectral index and relations above suggest that if slow-roll
inflation is right, then scalar and tensor spectra must both be
nearly scale invariant ($n_s\simeq 1$ and $n_t\simeq 0$).

\section{Temperature Anisotropies}

The primary goal of CMB experiments that map the temperature as
a function of position on the sky is recovery of the
temperature autocorrelation function or angular power spectrum
of the CMB.  The fractional temperature perturbation
$\Delta T(\hatn)/T$ in a given direction $\hatn$ can be expanded
in spherical harmonics,
\begin{equation}
     {\Delta T(\hatn) \over T} = \sum_{lm} \, a_{(lm)}^{\rm T}\,
     Y_{(lm)}(\hatn), \quad {\rm with} \quad     a_{(lm)}^{\rm
     T} = \int\, d\hatn\, Y_{(lm)}^*(\hatn) \, {\Delta T(\hatn)
     \over T}.
\label{eq:Texpansion}
\end{equation}
Statistical isotropy and homogeneity of the Universe imply that
these coefficients have expectation values $\VEV{ (a_{(lm)}^{\rm
T})^*
a_{(l'm')}^{\rm T}} = C_l^{\rm TT} \delta_{ll'} \delta_{mm'}$ when
averaged over the sky.  Roughly speaking, the multipole moments
$C_l^{\rm TT}$ measure the mean-square temperature difference
between two points separated by an angle $(\theta/1^\circ) \sim
200/l$.

\begin{figure}
\centerline{\psfig{file=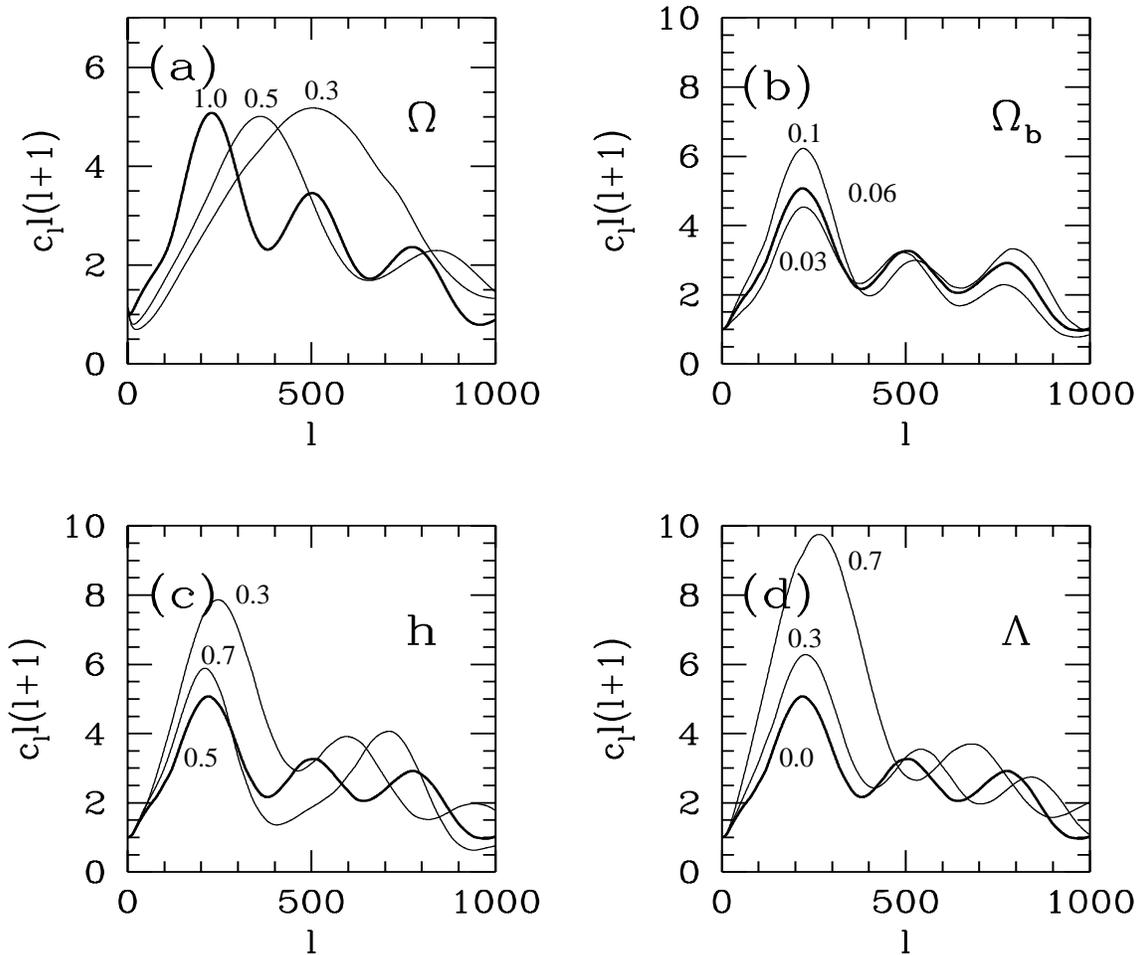,width=6in}}
\caption{
	  Theoretical predictions for CMB spectra as a function
	  of multipole moment $l$ for models with primordial
	  adiabatic perturbations.  In each case, the
	  heavy curve is that for the canonical standard-CDM values,
	  a total density $\Omega=1$, cosmological constant
	  $\Lambda=0$, baryon density $\Omega_b=0.06$, and
	  Hubble parameter $h=0.5$.  Each graph shows the effect
	  of variation of one of these parameters.  In (d),
	  $\Omega+\Lambda=1$.
}
\label{fig:models}
\end{figure}

Predictions for the $C_l$'s can be made given a theory for
structure formation and the values of several cosmological
parameters.  Fig.~\ref{fig:models} shows predictions for
models with primordial adiabatic perturbations.  The wriggles
come from oscillations in the photon-baryon fluid at the surface
of last scatter.  Each panel shows the effect of independent
variation of one of the cosmological parameters.  As
illustrated, the height, width, and spacing of the acoustic
peaks in the angular spectrum depend on  these (and other)
cosmological parameters.

These small-angle CMB anisotropies can be used to determine the
geometry of the Universe.\cite{kamspergelsug}  The angle
subtended by the horizon at the surface of last scatter is
$\theta_H \sim \Omega^{1/2} \;1^\circ$, and the peaks in the CMB
spectrum are due to causal processes at the surface of last
scatter.  Therefore, the angles (or values of $l$) at which the
peaks occur determine the geometry of the Universe.  This is
illustrated in Fig.~\ref{fig:models}(a) where the CMB spectra
for several values of $\Omega$ are shown.  As illustrated in the
other panels, the angular position of the first  peak is
relatively insensitive to the values of other undetermined (or
still imprecisely determined) cosmological parameters such as
the baryon density, the Hubble constant, and the cosmological
constant (as well as several others not shown such as the
spectral indices and amplitudes of the scalar and tensor spectra
and the ionization history of the Universe).  Therefore,
determination of the location of this first acoustic peak should
provide a robust measure of the geometry of the Universe.

\begin{figure}
\centerline{\psfig{file=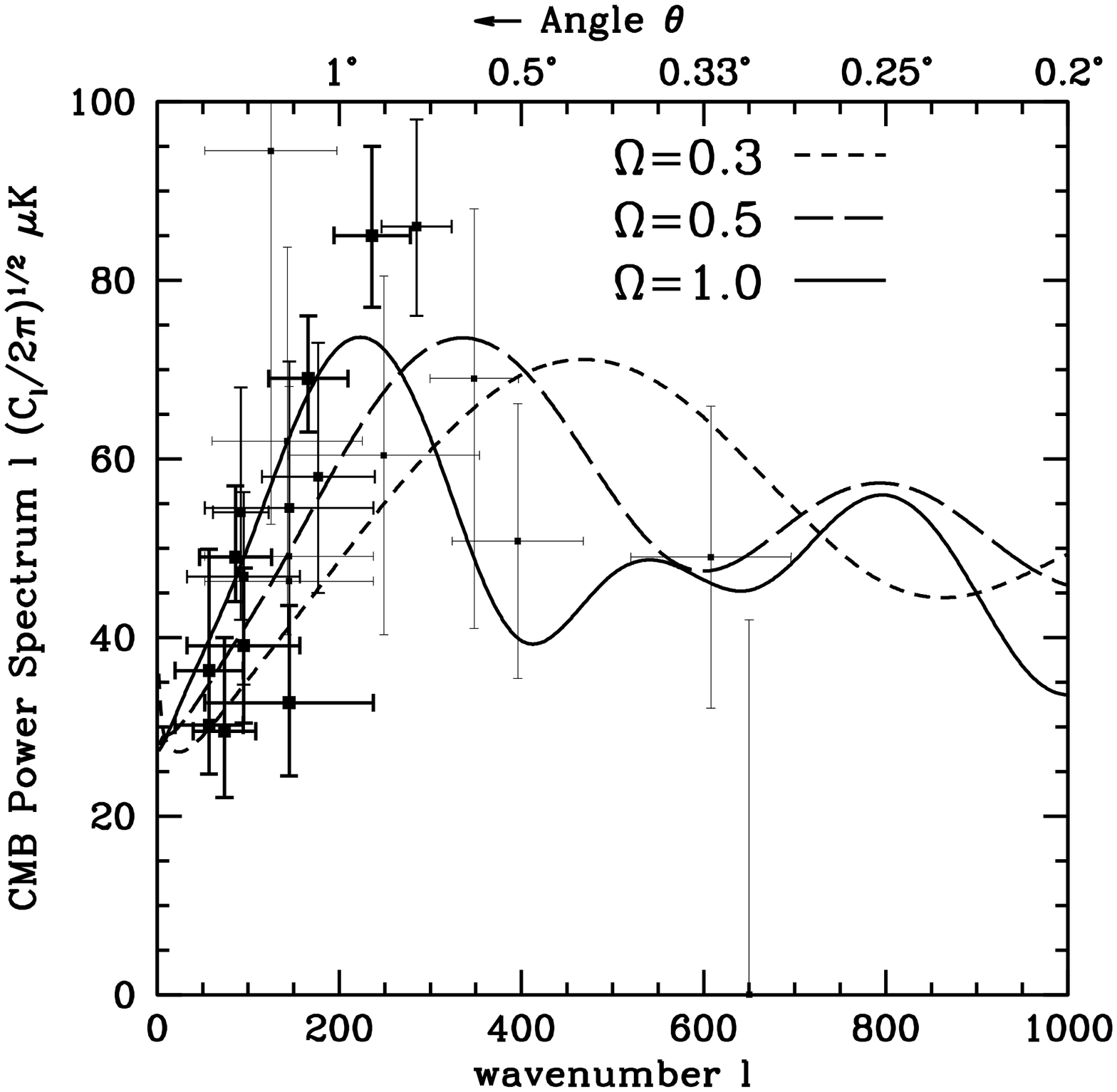,width=6in}}
\caption{Current CMB data.}
\label{fig:data}
\end{figure}

Fig.~\ref{fig:data} shows data from current ground-based and
balloon-borne experiments.  By fitting the theoretical curves to
these points, several groups find that the best fit to the data
is found with a total density $\Omega\simeq1.0$.\cite{current}
However, visual inspection of the data points in
Fig.~\ref{fig:data} clearly indicate that this current
determination of the geometry cannot be robust.

\begin{figure}
\centerline{\psfig{file=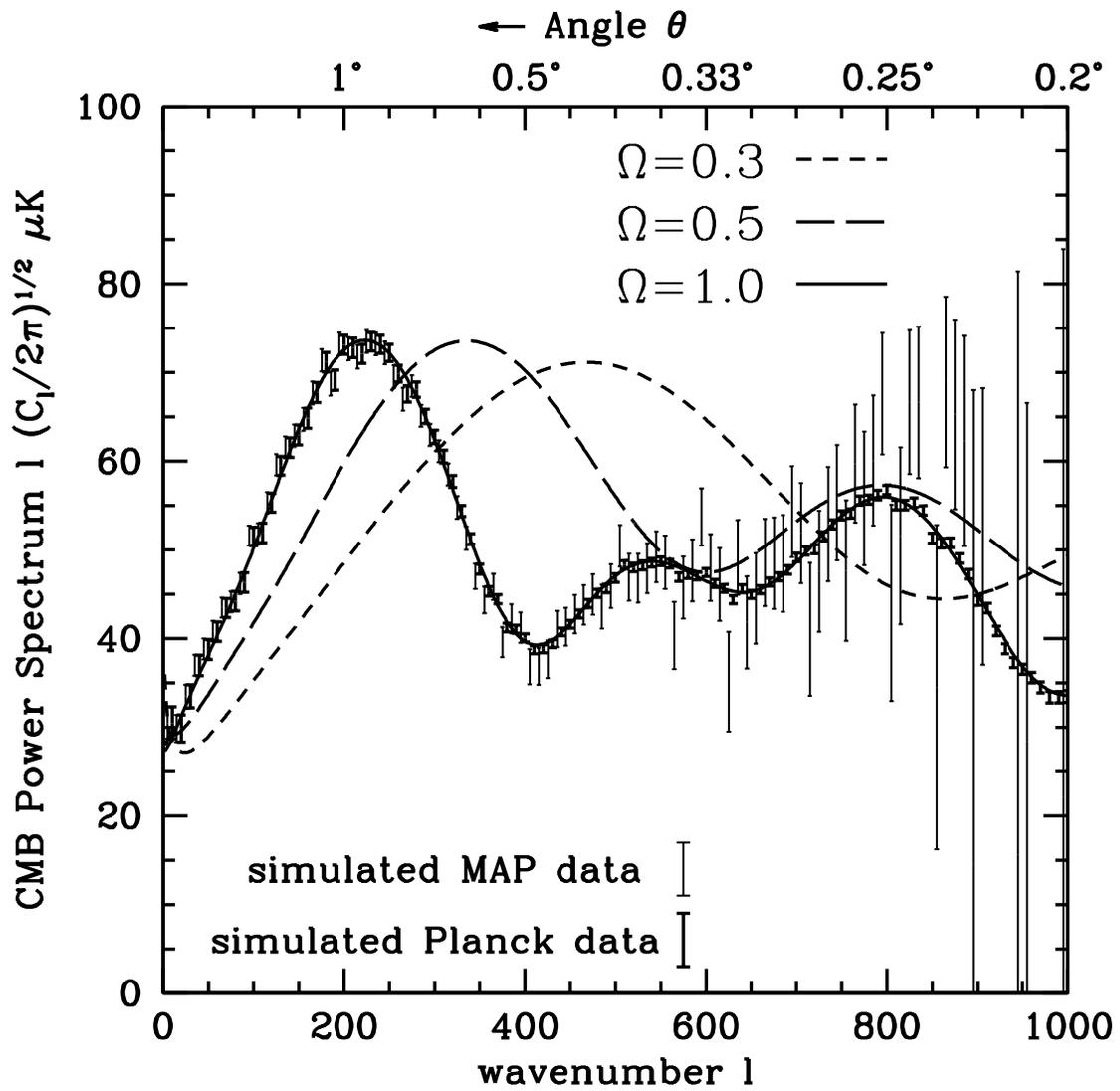,width=6in}}
\caption{Simulated MAP and Planck data.}
\label{fig:simulated}
\end{figure}

In the near future, the precision with which this determination
can be made will be improved dramatically.  NASA has recently
approved the flight of a satellite mission, the Microwave
Anisotropy Probe (MAP)\cite{MAP} in the year 2000 to carry out
these measurements, and ESA has approved the flight of a
subsequent more precise experiment, the Planck
Surveyor.\cite{PLANCK}  Fig.~\ref{fig:simulated} shows simulated
data from MAP and Planck.  The heavier points with smaller error bars
are those we might expect from Planck and the lighter points
with larger error bars are those anticipated for MAP.  Even
without any sophisticated analysis, it is clear from
Fig.~\ref{fig:simulated} that data from either of these
experiments will be able to locate the first acoustic peak
sufficiently well to discriminate between a flat Universe
($\Omega=1$) and an open Universe with $\Omega\simeq0.3-0.5$.

By doing what essentially boils down to a calculation of the
covariance matrix for such simulated data,
it can be shown that these satellite missions may
potentially determine $\Omega$ to better than 10\% {\it after}
marginalizing over all other undetermined parameters (we
considered 7 more parameters in addition to the 4 shown in
Fig.~\ref{fig:models}), and better than 1\% if the other parameters can be
fixed by independent observations or assumption.\cite{jkksone}
This would be far more accurate than any traditional
determinations of the geometry.

We also found that the CMB should provide values for
the cosmological constant and baryon density far more precise
than those from traditional observations.\cite{jkkstwo}  If there is more
nonrelativistic matter in the Universe than baryons can account
for---as suggested by current observations---it will become
increasingly clear with future CMB measurements.  Subsequent
analyses have confirmed these estimates with more accurate
numerical calculations.\cite{bet}

Although these forecasts relied on the assumptions that
adiabatic perturbations were responsible for structure formation
and that reionization would not erase CMB anisotropies, these
assumptions have become increasingly
justifiable in the past few years.  As discussed above,
the leading alternative theories for structure formation now
appear to be in trouble, and recent detections of CMB
anisotropy at degree angular separations show that the effects
of reionization are small.  

The predictions of a nearly scale-free spectrum of primordial
adiabatic perturbations will also be further tested with
measurements of small-angle CMB anisotropies.  The existence and
structure of the acoustic peaks shown in Fig.~\ref{fig:models}
will provide an unmistakable signature of adiabatic
perturbations\cite{huwhite} and the spectral index $n_s$ can be determined
from fitting the theoretical curves to the data in the same way
that the density, cosmological constant, baryon density, and
Hubble constant are also fit.\cite{jkkstwo}

Temperature anisotropies produced by a stochastic gravity-wave
background would affect the shape of the angular CMB spectrum,
but there is no way to disentangle the scalar and tensor
contributions to the CMB anisotropy in a model-independent way.
Unless the tensor signal is large, the cosmic variance from the
dominant scalar modes will provide an irreducible limit to the
sensitivity of a temperature map to a tensor signal.\cite{jkkstwo}

\section{CMB Polarization and Gravitational Waves}

Although a CMB temperature map cannot unambiguously distinguish
between the density-perturbation and gravity-wave contributions
to the CMB, the two can be decomposed in a model-independent
fashion with a map of the CMB
polarization.\cite{probe,ourpolarization,selzald}  Suppose we 
measure the linear-polarization ``vector'' $\vec P(\hatn)$ at
every point $\hatn$ on the sky.  Such a vector field can be written as the
gradient of a scalar function $A$ plus the curl of a vector
field $\vec B$: $\vec P(\hatn) \, = \, \vec \nabla A \, + \,
\vec\nabla \times \vec B.$
The gradient (i.e., curl-free) and curl components can be
decomposed by taking the divergence or curl of $\vec
P(\hatn)$ respectively.  Density perturbations are scalar metric
perturbations, so they have no handedness.  They can therefore
produce no curl.  On the other hand, gravitational waves {\it
do} have a handedness so they can (and we have shown that they
do) produce a curl.  This therefore provides a way to detect the
inflationary stochastic gravity-wave background and thereby
test the relations between the inflationary observables.  It
should also allow one to determine (or at least constrain in the
case of a nondetection) the height of the inflaton potential.

More precisely, the Stokes parameters $Q(\hatn)$ and $U(\hatn)$
(where $Q$ and $U$ are measured with respect to the polar ${\bf
\hat\theta}$ and azimuthal ${\bf \hat \phi}$ axes) which specify
the linear polarization in direction $\hatn$ are components of a
$2\times2$ symmetric trace-free (STF) tensor, 
\begin{equation}
  {\cal P}_{ab}(\hatn)={1\over 2} \left( \begin{array}{cc}
   \vphantom{1\over 2}Q(\hatn) & -U(\hatn) \sin\theta \\
   -U(\hatn)\sin\theta & -Q(\hatn)\sin^2\theta \\
   \end{array} \right),
\label{whatPis}
\end{equation}
where the subscripts $ab$ are tensor indices.
Just as the temperature is expanded in terms of spherical
harmonics, the polarization tensor can be expanded,\cite{ourpolarization}
\begin{equation}
      {{\cal P}_{ab}(\hatn)\over T_0} =
      \sum_{lm} \Biggl[ a_{(lm)}^{{\rm G}}Y_{(lm)ab}^{{\rm
      G}}(\hatn) +a_{(lm)}^{{\rm C}}Y_{(lm)ab}^{{\rm C}}(\hatn)
      \Biggr],
\label{Pexpansion}
\end{equation}
in terms of the tensor spherical harmonics $Y_{(lm)ab}^{\rm G}$
and $Y_{(lm)ab}^{\rm C}$, which are a complete basis for the
``gradient'' (i.e., curl-free) and ``curl'' components of the
tensor field, respectively.  The mode amplitudes are given by
\begin{equation}
a^{\rm G}_{(lm)}={1\over T_0}\int d\hatn\,{\cal P}_{ab}(\hatn)\, 
                                         Y_{(lm)}^{{\rm G}
					 \,ab\, *}(\hatn),
\qquad a^{\rm C}_{(lm)}={1\over T_0}\int d\hatn\,{\cal P}_{ab}(\hatn)\,
                                          Y_{(lm)}^{{\rm C} \,
					  ab\, *}(\hatn).
\label{Amplitudes}
\end{equation}
Here $T_0$ is the cosmological mean CMB temperature and $Q$ and
$U$ are given in brightness temperature units rather than flux
units.   Scalar perturbations have no handedness.  Therefore,
they can produce no curl, so $a_{(lm)}^{\rm C}=0$ for scalar
modes.  On the other hand tensor modes {\it do} have a
handedness, so they produce a non-zero curl, $a_{(lm)}^{\rm C}
\neq0$.

A given inflationary model predicts that the $a_{(lm)}^{\rm X}$
are gaussian random variables with zero mean,
$\VEV{a_{(lm)}^{\rm X}}=0$  (for ${\rm X},{\rm X}' = \{{\rm
T,G,C}\}$) and covariance $\VEV{\left(a_{(l'm')}^{\rm X'}
\right)^* a_{(lm)}^{\rm X}} = C_l^{{\rm XX}'}
\delta_{ll'}\delta_{mm'}$.   Parity demands that
$C_l^{\rm TC}=C_l^{\rm GC}=0$.  Therefore the statistics of the
CMB temperature-polarization map are completely specified by the
four sets of moments, $C_l^{\rm TT}$, $C_l^{\rm TG}$, $C_l^{\rm
GG}$, and $C_l^{\rm CC}$.   Also, as stated above, only tensor modes
will produce nonzero $C_l^{\rm CC}$.  

\begin{figure}
\centerline{\psfig{file=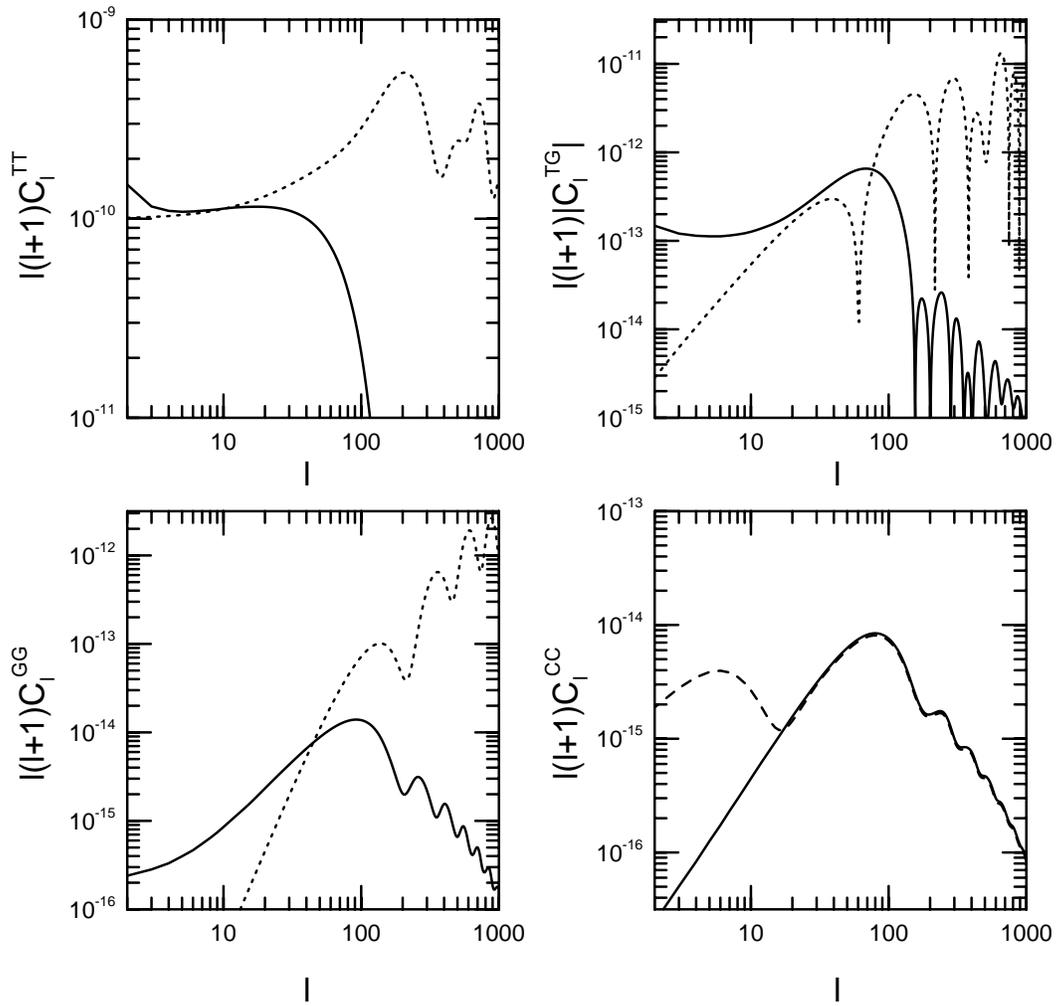,width=8in}}
\caption{
          Theoretical predictions for the four nonzero CMB
	  temperature-polarization spectra as a function
	  of multipole moment $l$.  The dashed line in the lower
	  right panel shows a reionized model with optical depth
	  $\tau=0.1$ to the surface of last scatter.
       }
\label{clsplot}
\end{figure}

To~illustrate, Fig.~\ref{clsplot} shows the four
temperature-polarization power spectra. The dotted curves
correspond to a COBE-normalized inflationary model
with cold dark matter and no cosmological constant
($\Lambda=0$), Hubble constant (in units of 100
km~sec$^{-1}$~Mpc$^{-1}$) $h=0.65$, baryon density
$\Omega_bh^2=0.024$, scalar spectral index $n_s=1$, no
reionization, and no gravitational waves.  The solid curves show
the spectra for a COBE-normalized stochastic gravity-wave
background with a flat scale-invariant spectrum ($h=0.65$,
$\Omega_b h^2=0.024$, and $\Lambda=0$) in a critical-density
Universe.   Note that the panel for $C_l^{\rm CC}$ contains no
dotted curve since scalar perturbations produce no C
polarization component.  The dashed curve in the CC panel shows
the tensor spectrum for a reionized model with optical depth
$\tau=0.1$ to the surface of last scatter.

As with a temperature map, the sensitivity of a polarization map
to gravity waves will be determined by the
instrumental noise and fraction of sky covered, and by the
angular resolution.  Suppose the detector sensitivity is $s$ and
the experiment lasts for $t_{\rm yr}$ years with an angular
resolution better than $1^\circ$.  Suppose further that we
consider only the CC component of the polarization in our
analysis.  Then the smallest tensor amplitude ${\cal T}_{\rm
min}$ to which the experiment will be sensitive at $1\sigma$
is\cite{detectability}
\begin{equation}
     {{\cal T}_{\rm min}\over 6\, C_2^{\rm TT}}
      \simeq 5\times 10^{-4} \left( {s\over \mu{\rm K}\,\sqrt{\rm
      sec}} \right)^2 t_{\rm yr}^{-1}.
\label{CCresult}
\end{equation}
Thus, the curl component of a full-sky polarization map is
sensitive to inflaton potentials $V\ga 5 \times
10^{-15}t_{\rm yr}^{-1}$ $(s/\mu{\rm K}\, \sqrt{\rm sec})^2$.  
Improvement on current constraints with only the curl
polarization component requires a detector sensitivity
$s\la40\,t_{\rm yr}^{1/2}\,\mu$K$\sqrt{\rm sec}$.  For
comparison, the detector sensitivity of MAP will be $s={\cal
O}(100\,\mu$K$\sqrt{\rm sec})$.  However, Planck may conceivably
get sensitivities around $s=25\,\mu$K$\sqrt{\rm sec}$.

\begin{figure}
\centerline{\psfig{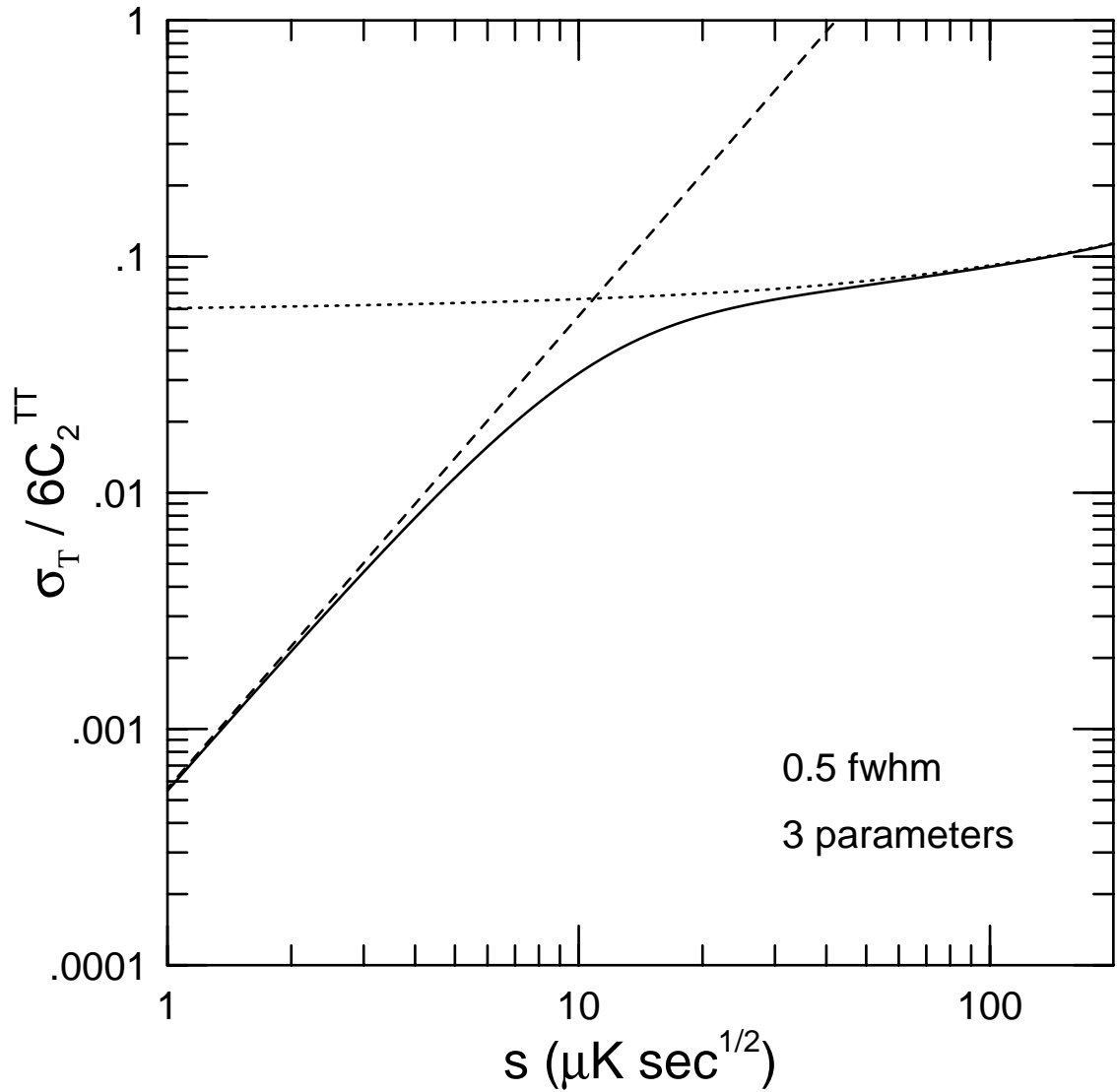}}
\caption{
         Results for the $1\sigma$ sensitivity $\sigma_{\cal T}$ to
	 the amplitude $\cal T$ of a flat ($n_t=0$) tensor spectrum
	 as a function of detector sensitivity $s$ for an
	 experiment which maps the CMB temperature and
	 polarization on the full sky for one year with an
	 angular resolution of $0.5^\circ$.  The vertical axis
	 is in units of the temperature quadrupole.  See text
	 for more details.}
\label{resultssimple}
\end{figure}

Eq.~\ref{CCresult} is the sensitivity obtained by using only
the curl component of the polarization, which provides a
model-independent probe of the tensor signal.  However, if we
are willing to consider specific models for the tensor and
scalar spectra, the sensitivity to a tensor signal may be
improved somewhat by considering the predictions for the full
temperature/polarization auto- and cross-correlation power
spectra.\cite{detectability}

For example, Fig.~\ref{resultssimple} shows the $1\sigma$
sensitivity $\sigma_{\cal T}$ to the amplitude $\cal T$ of a
flat ($n_t=0$) tensor spectrum as a function of detector
sensitivity $s$ for an experiment which maps the CMB temperature
and polarization on the full sky for one year with an angular
resolution of $0.5^\circ$.  The dotted curve
shows the results obtained by fitting only the TT moments; the
dashed curve shows results obtained by fitting only the CC
moments; and the solid curve shows results obtained by fitting
all four nonzero sets of moments. 

In Fig.~\ref{resultssimple}, we have assumed that the
spectra are fit only to $\cal S$, $\cal T$, and $n_s$, and the
parameters of the cosmological model are those used in
Fig.~\ref{resultssimple}.  If one fits to more cosmological
parameters (e.g., $\Omega_b$, $h$, $\Lambda$, etc.) as well, the
the sensitivity from the temperature moments, important for
larger $s$, is degraded.  However, the sensitivity due to the CC
component, which controls the total sensitivity for smaller $s$,
is essentially unchanged.  Again, this is because the CC signal
is very model-independent.

For detectors sensitivities $s\ga 20\,\mu$K$\,
\sqrt{\rm sec}$, the tensor-mode detectability for the
three-parameter fit shown in Fig.~\ref{resultssimple} comes
primarily from the temperature map, although polarization does
provides some incremental improvement.  However, if the data are
fit to more cosmological parameters (not shown), the
polarization improves the tensor sensitivity even for $s\ga
20\,\mu$K$\, \sqrt{\rm sec}$.  In any case, the
sensitivity to tensor modes comes almost entirely from the
curl of the polarization for detector sensitivities $s
\la10\,\mu$K$\sqrt{\rm sec}$.  Since the value of $s$ for Planck
will be somewhat higher, it will likely require a more
sensitive future experiment to truly capitalize on the
model-independent curl signature of tensor modes.

Finally, it should be noted that even a small amount of
reionization will significantly increase the polarization signal
at low $l$,\cite{reionization} as shown in the CC panel of
Fig.~\ref{clsplot} for $\tau=0.1$.  With such a level of
reionization (which may be expected in CDM models, as discussed
in the following Section), the sensitivity to the tensor amplitude is
increased by more than a factor of 5 over that in
Eq.~(\ref{CCresult}).  This level of reionization (if not more)
is expected in cold dark matter
models,\cite{kamspergelsug,blanchard,haiman} so if anything,
Eq.~(\ref{CCresult}) and Fig.~\ref{resultssimple} provide
conservative estimates.

\section{The Ostriker-Vishniac Effect and the Epoch of
Reionization}

Although most of the matter in CDM models does not undergo
gravitational collapse until relatively late in the history of
the Universe, some small fraction of the mass is expected to
collapse at early times.  Ionizing radiation released by this
early generation of star and/or galaxy formation will partially
reionize the Universe, and these ionized electrons
will re-scatter at least some cosmic microwave background (CMB)
photons after recombination
at a redshift of $z\simeq1100$.  Theoretical
uncertainties in the process of star formation and the resulting
ionization make precise predictions of the ionization history
difficult.  Constraints to the shape of the CMB blackbody
spectrum and detection of CMB anisotropy at degree angular
scales suggest that if reionization occurred, the fraction of
CMB photons that re-scattered is small.  Still, estimates show
that even if small, at least some reionization is expected in
CDM models:\cite{kamspergelsug,blanchard,haiman} for
example, the most careful recent calculations suggest a
fraction $\tau_r\sim0.1$ of CMB photons were re-scattered.\cite{haiman}

Scattering of CMB photons from ionized
clouds will lead to anisotropies at arcminute separations below
the Silk-damping scale (the Ostriker-Vishniac
effect).\cite{ostriker,andrewmarc}  These
anisotropies arise at higher order in perturbation theory and
are therefore not included in the usual Boltzmann calculations
of CMB anisotropy spectra.  The level of anisotropy is
expected to be small and it has so far eluded detection.  However,
these anisotropies may be observable with forthcoming CMB
interferometry experiments\cite{interferometers} that probe the
CMB power spectrum at arcminute scales.

\begin{figure}
\centerline{\psfig{file=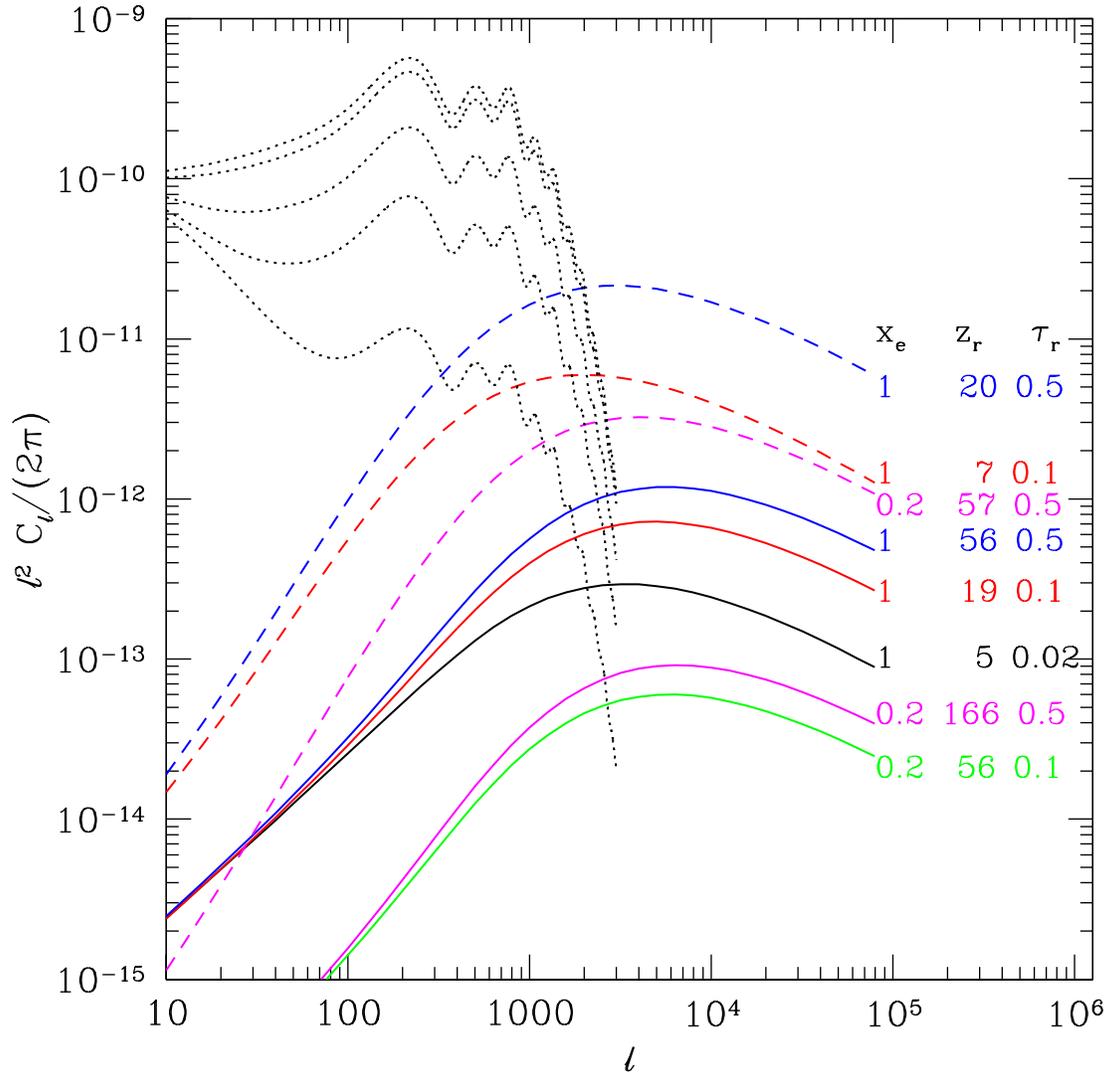,width=6in}}
\caption{Multipole moments for the Ostriker-Vishniac effect for
     the COBE-normalized canonical standard-CDM model
     ($\Omega=1$, $h=0.5$, $n=1$, $\Omega_b h^2 =0.0125$), for a
     variety of ionization histories.as listed. 
     We also show predictions for several open
     high-baryon-density models with the same $x_e$ and
     $\tau_r$, normalized to the cluster abundance, with dashed
     curves.  The dotted curves show the primary anisotropy for
     this model for $\tau_r=0.0$, 0.1, 0.5, 1, and 2, from top
     to bottom.
}
\label{fig:Clsplot}
\end{figure}

Fig.~\ref{fig:Clsplot}~\cite{andrewmarc} shows the predicted
temperature-anisotropy spectrum from the Ostriker-Vishniac
effect for a number of ionization histories.  The
ionization histories are parameterized by an ionization fraction
$x_e$ and a redshift $z_r$ at which the Universe becomes
reionized.  The optical depth $\tau$ to the
standard-recombination surface of last scatter can be obtained
from these two parameters.  

Reionization damps the acoustic peaks in the
primary-anisotropy spectrum by $e^{-2\tau}$, as shown in
Fig.~\ref{fig:Clsplot}, but this damping is essentially
independent of the details of the ionization history.  That is,
any combination of $x_e$ and $z_r$ that gives the same $\tau$
has the same effect on the primary anisotropies.  So although
MAP and Planck will be able to determine $\tau$ from this
damping, they will not constrain the epoch of reionization.  On
the other hand, the secondary anisotropies (the
Ostriker-Vishniac anisotropies) produced at smaller angular
scales in reionized models {\it do} depend on the ionization
history.  For example, although the top and bottom dashed curves
in Fig.~\ref{fig:Clsplot} both have the same optical depth, they
have different reionization redshifts ($z_r=20$ and $z_r=57$).
Therefore, if MAP and Planck determine $\tau$, the amplitude of
the Ostriker-Vishniac anisotropy determines the reionization
epoch.\cite{andrewmarc}

In a flat Universe, CDM models  normalized to cluster abundances
produce rms temperature anisotropies of 0.8--2.4 $\mu$K on
arcminute angular scales for a constant ionization fraction of
unity, whereas an ionization fraction of 0.2 yields rms
anisotropies of 0.3--0.8 $\mu$K.\cite{andrewmarc}  In an open and/or
high-baryon-density Universe, the level of anisotropy is
somewhat higher.  The signal in some of these models may be
detectable with planned interferometry experiments.\cite{andrewmarc}

\section{Discussion}

If MAP and Planck find a CMB temperature-anisotropy spectrum
consistent with a flat Universe and nearly--scale-free
primordial adiabatic perturbations, then the next step will be
to isolate the gravity waves with the polarization of the CMB.
If inflation has something to do with grand unification, then it
is possible that Planck's polarization sensitivity will be
sufficient to see the polarization signature of gravity waves.
However, it is also quite plausible that the height of the
inflaton potential may be low enough to elude detection by
Planck.  If so, then a subsequent experiment with better
sensitivity to polarization will need to be done.

Inflation also predicts that the distribution of primordial
density perturbations is gaussian, and this can be tested with
CMB temperature maps and with the study of the large-scale
distribution of galaxies.  Since big-bang nucleosynthesis
predicts that the baryon density is $\Omega_b \la 0.1$ and
inflation predicts $\Omega=1$, another prediction of inflation
is a significant component of nonbaryonic dark matter.  This can
be either in the form of vacuum energy (i.e., a cosmological
constant), and/or some new elementary particle.  Therefore,
discovery of particle dark matter could be interpreted as
evidence for inflation.

Large-scale galaxy surveys will soon map the distribution of
mass in the Universe today, and CMB experiments will shortly
determine the mass distribution in the early Universe.  The next
step will be to fill in the precise history of structure
formation in the ``dark ages'' after recombination but before
redshifts of a few.  Reconstruction of this epoch of cosmic
history will likely require amalgamation of the complementary
information provided by a number observations in several
wavebands.  Detection of
the secondary CMB anisotropies at arcminute scales produced by
scattering from reionized clouds will provide an indication of
the epoch of reionization, and therefore the epoch at which
structures first undergo gravitational collapse in the
Universe.\cite{andrewmarc}

\section*{Acknowledgments}
This work was supported by the U.S. D.O.E. under contract
DEFG02-92-ER 40699, NASA NAG5-3091, and the Alfred P. Sloan
Foundation at Columbia.

\end{document}